\documentclass{aa}  

\usepackage{graphicx}
\usepackage{txfonts}
\usepackage{hyperref}
\usepackage{mathtools}
\usepackage{newtxtext}
\usepackage{newtxmath}
\usepackage{subfig}
\usepackage{blindtext}
\usepackage{tabularx,ragged2e}
\usepackage{color} 
\usepackage{longtable}
\usepackage[dvipsnames]{xcolor}

\emergencystretch=\maxdimen
\newcommand{\tabhead}[1]{\textbf{#1}}
\usepackage{subcaption} 

\begin{document}
	
    \title{High-precision broadband linear polarimetry of early-type binaries. V. The Binary System HD 165052 in Open Cluster NGC 6530\thanks{The polarization data for HD 165052 are only available in electronic form at the CDS via anonymous ftp to \url{cdsarc.cds.unistra.fr} (\url{130.79.128.5}) or via \url{https://cdsarc.cds.unistra.fr/cgi-bin/qcat?J/A+A/}}}
	
    \titlerunning{HD~165052}
	
    \author{Yasir Abdul~Qadir\inst{1}
    \and  Andrei V. Berdyugin\inst{1}
    \and Vilppu Piirola\inst{1}
    \and Takeshi Sakanoi\inst{2} \and \\
    Masato Kagitani\inst{2} 
    \and Svetlana V. Berdyugina\inst{3,4,5}
    }
	
    \institute{Department of Physics and Astronomy, FI-20014 University of Turku, Finland \\
    \email{yasir.abdulqadir@utu.fi}
    \and Graduate School of Sciences, Tohoku University, Aoba-ku, 980-8578 Sendai, Japan 
    \and Istituto ricerche solari Aldo e Cele Dacc\'o (IRSOL), Faculty of Informatics, Universit\`a della Svizzera italiana, Via Patocchi 57, Locarno, Switzerland
    \and Euler Institute, Faculty of Informatics, Universit\`a della Svizzera italiana, Via la Santa 1, 6962 Lugano, Switzerland
    \and Institut f\"ur Sonnenphysik (KIS), Georges-K\"ohler-Allee 401A, 79110 Freiburg, Germany\\
    }
	
    \date{Received XX / Accepted XX}
    
    \abstract{}{This is a continuation of our study of early-type binaries with high-precision broad-band polarimetry. This time, we are presenting results of our observations of the massive O+O-type binary system HD 165052 located in the very young open cluster NGC 6530. Our aim was to investigate the variations of linear polarization in this system with time and obtain independent estimate of orbital period from polarization data. By fitting phase-locked variations of Stokes parameters $q$ and $u$, it is possible to obtain independent estimates of orbit inclination $i$, orientation $\Omega$, and the direction of rotation.} 
    {We employed the Dipol-2 polarimeter in combination with the remotely controlled 60cm KVA and Tohoku T60 telescopes. Linear polarization measurements of HD 165052 were obtained in the $B$, $V$, and $R$ passbands with an accuracy better than $\sim$$0.01\%$. Lomb-Scargle method of period search was applied to the acquired polarization data to identify present periodic signals. To study interstellar polarization in NGC 6530, we observed 25 cluster stars located in close proximity to the binary.}
    {Our observations clearly revealed periodic variations of Stokes parameters in all three passbands with the amplitude $\sim0.1\%$. Period search, applied to polarization data, detected unambiguous and strongest periodic signal at 1.47755 $\pm$ 0.005 d, that corresponds to half of the orbital period of 2.95510 $\pm$ 0.005 d. Our independent period search, performed on polarization data, supports a "shorter" value of the orbital period in HD 165052 determined from RV measurements done in the past and rejects "longer" value obtained very recently. In the observed polarization variations, the second terms of the Fourier harmonics are clearly dominating suggesting that light scattering material is symmetrically distributed with respect to the orbital plane. We concluded that the most probable polarization mechanism is the electron scattering in the interacting stellar winds. To evaluate the effect of non-periodic noise on the orbital parameters derived from analysis of periodic variations of polarization, we have employed rigorous Monte Carlo simulations for the available data. Thus, we derived our best estimate of the orbital inclination $i$ as 55$^{\circ} + 5^{\circ}/-55^{\circ}$, and $\Omega$ as 148$^{\circ} (328^{\circ}) + 20^{\circ}/-22^{\circ}$ averaged over $B$, $V$, and $R$ passbands. Our estimates provide direct and unambiguous evidence that non-periodic noise, if present, significantly reduce accuracy of orbital parameters obtained from high-precision polarization data for the low-inclination systems. Using the values of polarization variability amplitude, we estimated the mass-loss rate for the whole system to be \( \dot{M} = 3.47 \times 10^{-7} \pm 1.59 \times 10^{-8}M_{\rm \odot} \; \rm yr^{-1} \). The direction of the binary system rotation on the plane of the sky is clockwise. Our observations of HD 165052 neighbour stars revealed complex behaviour of interstellar polarization in the cluster NGC 6530.}{}

    \keywords{polarization  -- techniques: polarimetric   -- instrumentation: polarimteters -- stars: individual: HD~165052 -- binaries (including multiple): close -- binaries: non-eclipsing}
	
    \maketitle
    \section{Introduction}\label{sec:intro}

    \begin{figure*}[htp]
    \centering
    \includegraphics[width=1\textwidth]{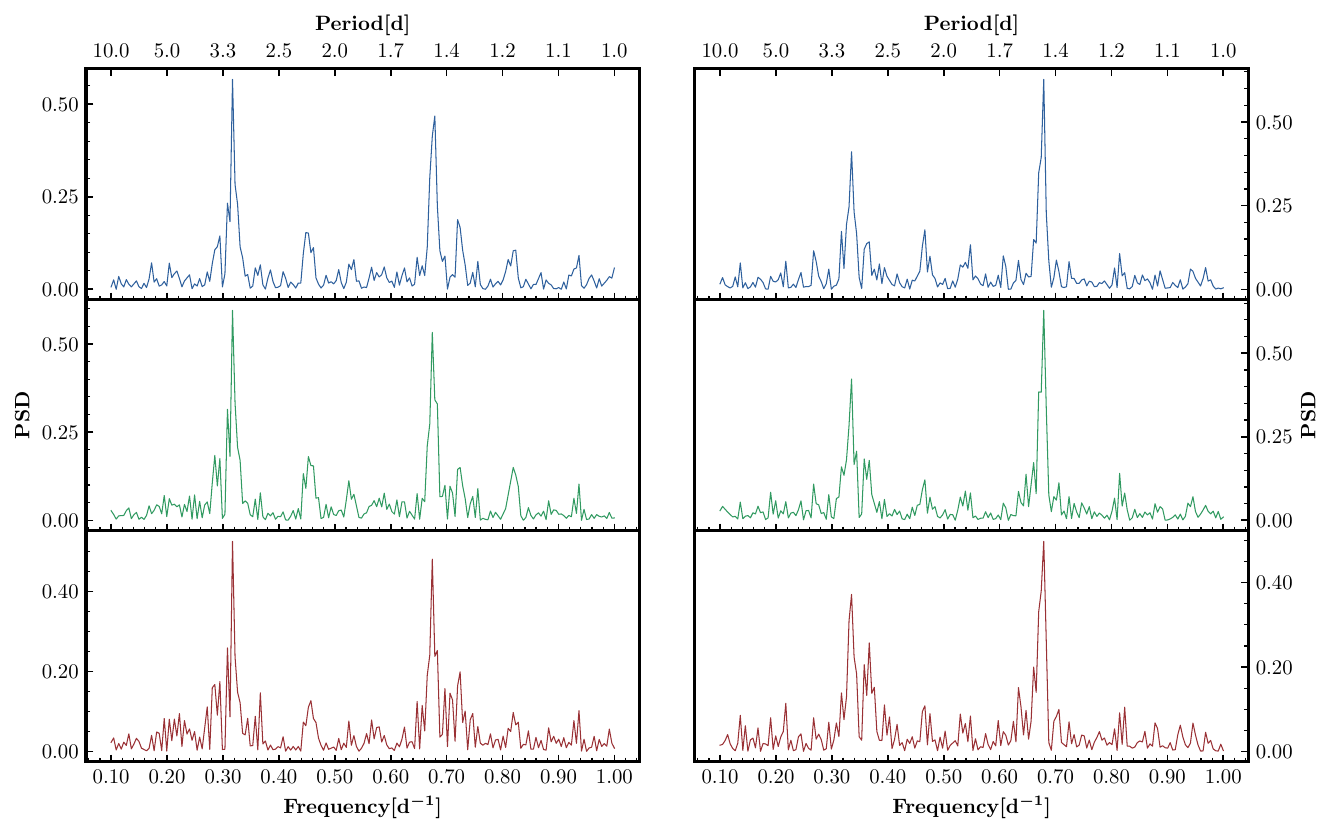}
    \caption{Lomb-Scargle periodograms for Stokes $q$ (left column) and $u$ (right column) for HD~165052 obtained in the $B$, $V$, and $R$ passbands (top, middle, and bottom panels respectively).} 
    \label{fig:ls}
    \end{figure*}

    Early-type binaries are massive and luminous binary stars with both companions of early B, or O spectral types. One of the key features of such binaries is that they often exhibit a noticeable amount of circum-binary material even if neither of the companions is filling its Roche lobe. This is occurring due to the extensive stellar winds. Interaction between these winds results in formation of hot bow shock area of the ionized gas between and around the companions. Light scattering on such gas may give rise to the variable linear polarization. Studying such polarization provides important information on the orbital parameters of the system and geometry / distribution of the circum-binary material. We have been conducting polarimetric observations of early-type binaries using DiPol-2 polarimeter for about a decade from now. In our previous papers  we have described results obtained for early-type binaries HD~48099 \citep{2016A&A...591A..92B}, AO Cassiopeiae \citep{2023A&A...670A.176A}, and DH Cephei \citep{2023A&A...677A..75A}. 
    In this paper we are presenting results of our polarimetric study of the early-type binary system HD~165052. 

    HD 165052 is a double-lined spectroscopic binary that consists of O7V$z$ primary star and O7.5V$z$ secondary star with a mass ratio $q = M_{\rm s}/M_{\rm p} = 0.91 \pm 0.01$ \citep{2013MNRAS.433.1300F}.  The orbital period value determined in last $\sim20$ years ranges from 2.95510d \citep{2002MNRAS.333..202A} to 2.95585d \citet{2023MNRAS.521.2988R}. Due to the noticeable eccentricity of orbit ($e=0.09$) and clear presence of the apsidal motion in HD 165052, a number of detailed spectroscopic investigations have been conducted for this system in the past (e.g., \citet{1978ApJ...224..558M}; \citet{1997Obs...117..295S}; \citet{2013MNRAS.433.1300F}; \citet{2002MNRAS.333..202A}; \citet{2007A&A...474..193L}; \citet{2023MNRAS.521.2988R}). HD 165052 belongs to the very young open cluster NGC 6530, which is a well studied cluster of our Galaxy because of its interesting history of star formation (e.g., \citet{2000AJ....120..333S}; \citet{2006A&A...459..477D}; \citet{2020PASP..132d4301T}).

    As far as we know, no comprehensive polarization studies of HD 165052 and NGC 6530 have been undertaken in the past. Therefore, we carried out a series of precise broadband linear polarization observations of HD 165052 and measured linear polarization of its 24 nearby stars within NGC 6530 cluster. Our main goal was to detect potential orbital phase-locked variable linear polarization in the HD 165052. This allows to obtain independent estimates of its orbital parameters and study the distribution of circum-binary material. We emphasize that this goal can be achieved even if the exact amount of interstellar polarization component cannot be determined.

    \begin{table}
    \caption{FAPs for the real and alias periods obtained for HD 165052 from period search.} 
    \label{table:fap}
    \centering
    \begin{tabular}[c]{l c c c} 
    \hline\hline 
    & Filter & Period[d] & FAP \\ \hline
    
    Stokes $q$: & $B$ & 1.47755 & $8.5 \times 10^{-24}$ \\  
    && 3.07101 & $3.9 \times 10^{-18}$ \\  
    & $V$ & 1.47756 & $6.7 \times 10^{-22}$ \\  
    && 3.07096 & $2.8 \times 10^{-18}$ \\  
    & $R$ & 1.47757 & $1.7 \times 10^{-18}$ \\
    && 3.07093 & $9.3 \times 10^{-16}$ \\  

    Stokes $u$: & $B$ & 1.47754 & $2.5 \times 10^{-24}$ \\  
    && 3.07108 & $2.3 \times 10^{-18}$ \\  
    & $V$ & 1.47755 & $2.3 \times 10^{-19}$ \\  
    && 3.07105 & $1.4 \times 10^{-14}$ \\  
    & $R$ & 1.47754 & $4.5 \times 10^{-19}$ \\
    && 3.07105 & $4.8 \times 10^{-15}$ \\  

    \hline
    \end{tabular}
    \end{table}

    \section{Polarimetric Observations}\label{sec:observations}
    First polarization data set spanning over 32 nights was obtained in 2012 with the DiPol-2 \citep{2014SPIE.9147E..8IP} polarimeter on the 60cm KVA telescope at the Observatory Roque de los Muchachos (La Palma, Canary Islands). In 2015 and 2016, we gathered another sets of data that spanned over 9 nights and 4 nights subsequently with the same polarimeter on the 60cm Tohoku T60 telescope at Haleakal\={a} Observatory (Hawaii). Finally, a last set of data, spanning over 23 nights, was acquired with the T60 telescope in 2023. In the year 2023, 24 field stars, members of NGC 6530 cluster in the vicinity of HD 165052 were also observed. All our polarimetric observations were conducted in the $B$, $V$, and $R$ passbands simultaneously in remote mode. For HD~165052, we usually obtained 256 images with an exposure time of 10~s per each night, which yielded 64 measurements of the Stokes $q$ and $u$ parameters. Corresponding observational log is given in Table \ref{table:log}. 

    The process of data reduction involved standard calibration, including bias and dark frame subtraction. Correction for flat-field effects is also done, although this correction does not impact final results significantly. Normalized Stokes $q$ and $u$ parameters were derived from flux intensity ratios of orthogonally polarized stellar images $Q_{\rm i} = I_{\rm e}(i)/I_{\rm o}(i)$, obtained at orientations of half-wave plate $i = 0.0^\circ$, $22.5^\circ$, $45.0^\circ$, $67.5^\circ$,.., $337.5^\circ$. More details can be found in \citet{2020A&A...635A..46P} and \citet{2023A&A...670A.176A}.

    To determine instrumental polarization, we observed about two dozen non-polarized standard stars. The instrumental polarization for both telescopes in the $B$, $V$, and $R$ bands ranged from $0.004\%$ to $0.006\%$, determined with the accuracy of $0.0002\%$ -- $0.0003\%$. Highly polarized standard stars HD 204827 and HD 25443 were used to calibrate the polarization angle zero-point (see Table \ref{table:hp} for their polarization degrees and angles). The errors of the nightly average polarization measurements for T60 are from $0.001\%$ to $0.004\%$ in the $B$ and $R$ bands and from $0.002\%$ to $0.006\%$ in the $V$ band. For the data obtained with the KVA, the errors are somewhat larger, and are in the range from $0.006\%$ to $0.015\%$.

    \section{Data Analysis}\label{sec:analysis}
    \subsection{Period Search}\label{sec:prdsch} 

    To search for periodic signals in our polarimetric data, we employed Lomb-Scargle algorithm \citep{1976Ap&SS..39..447L, 1982ApJ...263..835S}. The benefit of using Lomb-Scargle algorithm is that it uses the least squares method to fit sinusoidal on unevenly sampled data, which it is often the case with astronomical data. To execute Lomb-Scargle algorithm with \textsc{Python} code, we used package from \texttt{astropy.timeseries}\footnote{\url{https://docs.astropy.org/en/stable/timeseries/ lombscargle.html}} \citep{2018zndo...1211397P}. 

    The variations of Stokes parameters of polarization arising from the scattered light in the binary system with (near) circular orbit typically show two maxima and minima per orbital period separated by 0.5 in the orbital phase \citep{1978A&A....68..415B}. Therefore, Lomb-Scargle algorithm applied to binary polarization data, is expected to detect {\it half} of the orbital period. We have plotted Lomb-Scargle periodograms for both Stokes  $q$ and $u$ in $B$, $V$, and $R$ passbands in Figure \ref{fig:ls}. 
    
   As it was expected, Lomb-Scargle algorithm has detected half of the orbital period, $P_{\text{orb/2}}$, with a well-constrained error of $\sigma_{\text{orb/2}}$. While the full orbital period, $P_{\text{orb}} = 2 \times P_{\text{orb/2}}$, the errors on $P_{\text{orb}}$ should remain equivalent to that of $P_{\text{orb/2}}$, i.e., $\sigma_{\text{orb}} = \sigma_{\text{orb/2}}$. 
    For Stokes $q$, we found the best orbital periods as $P_{\text{orb}}$ = 2.95510 $\pm$ 0.006~d, 2.95512 $\pm$ 0.007~d, and 2.95514 $\pm$ 0.008~d in the $B$, $V$, and $R$ passbands respectively. Likewise, for Stokes $u$, we found the best periods as $P_{\text{orb}}$ = 2.95508 $\pm$ 0.006~d, 2.95510 $\pm$ 0.007~d, and 2.95508 $\pm$ 0.008~d. Thus, the derived average value over Stokes $q$ and $u$ and across all passbands is $P_{\text{orb}}$ = 2.95510 $\pm$ 0.005~d. The errors on the periods were calculated using corresponding power values and measurement uncertainties of the frequencies related to orbital period peaks given in Table \ref{table:fap}. These frequency errors were then converted to days using the relationship $\Delta P = \Delta f/ f^2$.    
    
    Moreover, we do notice the presence of strong alias peak at $\sim$$3.1$~d (frequency $\sim$$0.32~$d$^{-1}$) in all periodograms. An alias peak related to real period can appear when the frequency of the that period is not less than half of the sampling frequency, i.e., Nyquist frequency ($f_{\rm ny}$). In such a situation, two waves are produced that differ by 1/$f_{\rm ny}$. The (half) orbital period for HD 165052, $\sim$1.47755~d is at frequency of $\sim$0.68$~$d$^{-1}$, which is more than half of our sampling frequency. Therefore, alias peaks can be expected at around 1/(1-1/1.47755) = 3.094~d or at frequency close to $\sim$0.323~d$^{-1}$. As one can see, the alias peaks which are quite close to that frequency indeed present in our periodograms (cf. \citet{1949IEEEP..37...10S}; \citet{2018ApJS..236...16V}). It is not uncommon to observe such alias peaks in Lomb-Scargle periodograms of astronomical data (e.g., \citet{2018MNRAS.478.4710K}; \citet{2023A&A...670A.176A}; \citet{2023A&A...677A..75A}). 

    \begin{figure}[]
    \centering
    \includegraphics[width=\linewidth]{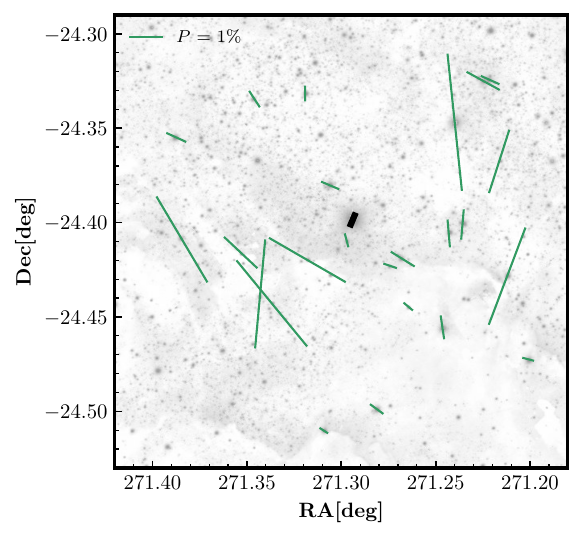}
    \caption{Interstellar polarization map in the vicinity of HD~165052. Measurements made in the $V$ passband are presented. Binary is in the center of the field and its average value of polarization  shown with thicker black line. The length of bars corresponds to the degree of linear polarization $P$, and the direction corresponds to the polarization angle (measured from the north to the east).} 
    \label{fig:map}
    \end{figure}

    Furthermore, we calculated False Alarm Probability (FAP) for (half) orbital period of HD 165052, and it's alias peaks using bootstrap method \citep{2012ada..confE..16S}. FAP values of both real and alis peaks in $B$, $V$, and $R$ passbands for both Stokes $q$, and $u$ are given in Table \ref{table:fap}. The FAP values for both orbital peaks and alias peaks are very low, because the algorithm can not distinguish between them. The value of true orbital period derived from our polarization data is very close to the values of 2.95510d derived by \citet{2002MNRAS.333..202A} and 2.95515 d derived by \citet{2007A&A...474..193L}. As we have found, the longer value of orbital period 2.95585 d, obtained recently by \citet{2023MNRAS.521.2988R} cannot be used to provide an adequate fit to observed variability of our polarization data. Thus, our independent period search, performed on polarization data, provides support for the previously measured shorter value of the orbital period in HD 165052. We did not detect presence of any other (real) periodic signal in polarimetric data except for the orbital.

    \begin{figure*}
    \centering
    \includegraphics[width=1\textwidth]{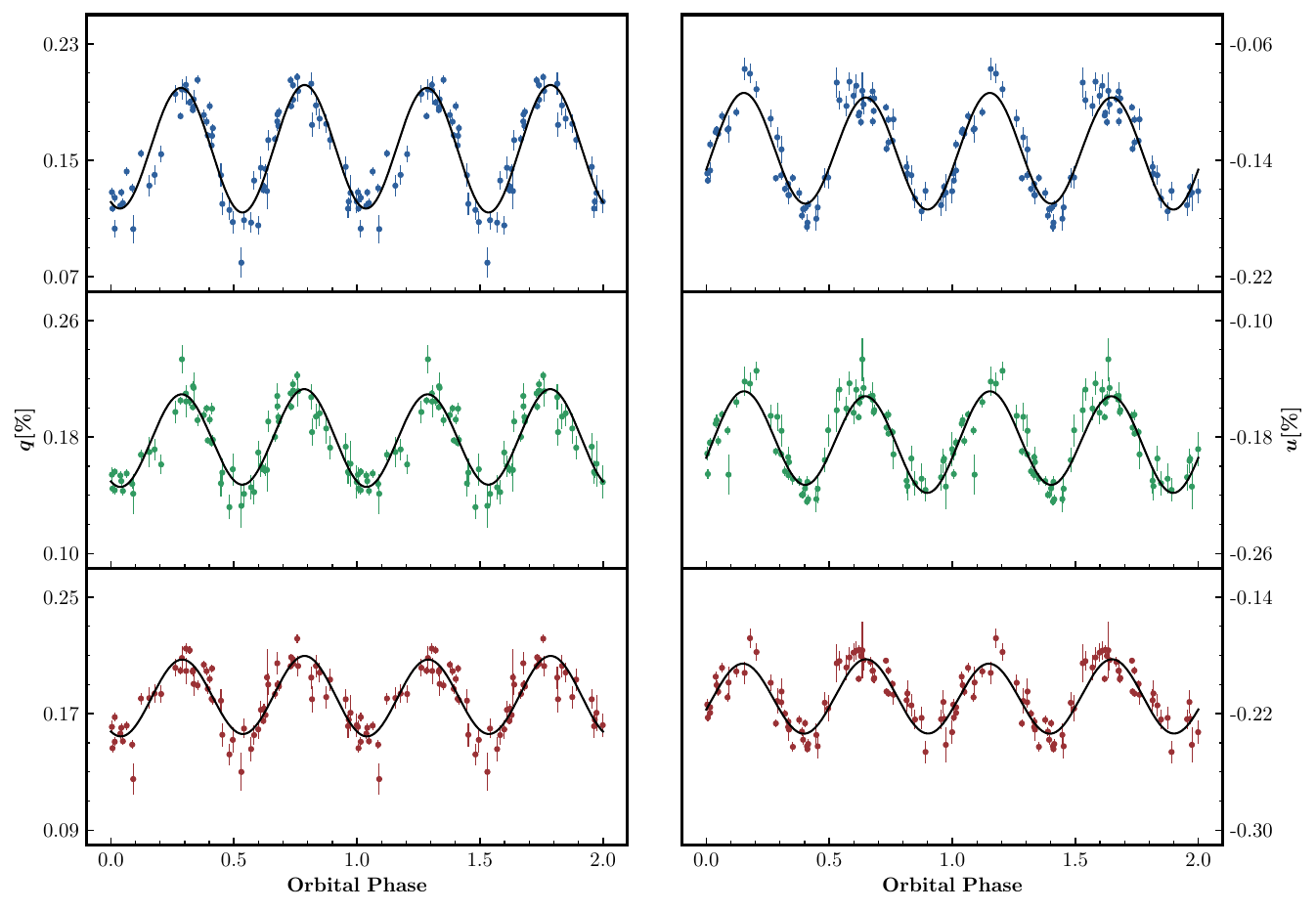}
    \caption{The variability of observed Stokes parameters $q$ and $u$ for HD~165052 in the $B$, $V$, and $R$ passbands (top, middle, and bottom panels respectively), phase-folded with the orbital period of 2.95510d. Fourier fit curves (see Sect. \ref{sec:fcm}) are shown with solid lines, and the best fit Fourier coefficients are given in Table \ref{table:fc}. The lengths of the bars correspond to $\pm\sigma$ errors.} 
    \label{fig:fs}
    \end{figure*}

    \begin{table*}
    \centering
    \caption{Best fit Fourier coefficients for Stokes $q$ and $u$. }
    \label{table:fc}
    \centering
    \scalebox{1.0}{
    \begin{tabular}[c]{ l c c c c c c c c c c }
    \hline\hline 
    \tabhead{\shortstack{\rm Filter }} & \tabhead{\shortstack{\rm $q_0$}}  & \tabhead{\shortstack{\rm $q_1$}} 
    & \tabhead{\shortstack{\rm $q_2$}} & \tabhead{\shortstack{\rm $q_3$}} & \tabhead{\shortstack{\rm $q_4$}}
    & \tabhead{\shortstack{\rm $u_0$}} & \tabhead{\shortstack{\rm $u_1$}} & \tabhead{\shortstack{\rm $u_2$}}
    & \tabhead{\shortstack{\rm $u_3$}} & \tabhead{\shortstack{\rm $u_4$}} \\

    \hline
    $B$ & 0.1581 & 0.0015 & -0.0007 & -0.0384 & -0.0185
    & -0.1335 & -0.0007 & 0.0025 & -0.0121 & 0.0363 \\
    & $\pm$0.0016 & $\pm$0.0021 & $\pm$0.0028 & $\pm$0.0022 & $\pm$0.0023
    & $\pm$0.0013 & $\pm$0.0022 & $\pm$0.0021 & $\pm$0.0022 & $\pm$0.0020 \\
				
    $V$ & 0.1788 & -0.0004 & -0.0019 & -0.0289 & -0.0143
    & -0.1829 & -0.0012 & 0.0030 & -0.0101 & 0.0310 \\
    & $\pm$0.0012 & $\pm$0.0017 & $\pm$0.0021 & $\pm$0.0020 & $\pm$0.0020 
    & $\pm$0.0011 & $\pm$0.0019 & $\pm$0.0017 & $\pm$0.0018 & $\pm$0.0017 \\	

    $R$ & 0.1819 & -0.0005 & -0.0014 & -0.0235 & -0.0123
    & -0.2088 & -0.0008 & -0.0012 & -0.0074 & 0.0236 \\
    & $\pm$0.0010 & $\pm$0.0015 & $\pm$0.0018 & $\pm$0.0016 & $\pm$0.0016 
    & $\pm$0.0012 & $\pm$0.0018 & $\pm$0.0019 & $\pm$0.0017 & $\pm$0.0018 \\				
    \hline	
    \end{tabular}}
    \end{table*}

    \subsection{Interstellar Polarization}\label{sec:isp}

    HD 165052, being a rather distant binary system with a parallax of $0.7893\pm0.0297$~mas \citep{2021A&A...649A...1G}, is expected to exhibit interstellar (IS) polarization component due to the presence of interstellar dust along the line of sight. In addition, one can expect a significant contribution to the IS polarization from the dust which presents in the young stellar cluster NGC 6530. We have attempted to quantify the IS component of polarization in HD 165052 by conducting polarization observations of 24 field stars located in the vicinity of HD 165052. Results of these observations are presented in Table \ref{table:is}. For each star, we give its Gaia DR3 identifier \citep{2021A&A...649A...1G}, coordinates, parallax, polarization degree, polarization angle, and the total number of polarization measurements in the three passbands.

    Figure \ref{fig:map} illustrates the polarization map of NGC 6530, showing a significant scatter in both polarization degrees (ranging from 0.07\% to 4.15\%) and polarization angles (ranging from $2^{\circ}$ to $176^{\circ}$) for the observed field stars, despite the fact that they all have parallaxes within the range of $\sim$0.3~mas. This variation in polarization properties among the cluster stars, including those closest to HD 165052, indicates that the IS polarization in NGC 6530 is strongly not uniform. Consequently, polarization data for the observed field stars cannot help us in accurate determination of IS polarization component for HD 165052 itself.

    \begin{figure*}
    \centering
    {\includegraphics[width=1.0\textwidth]{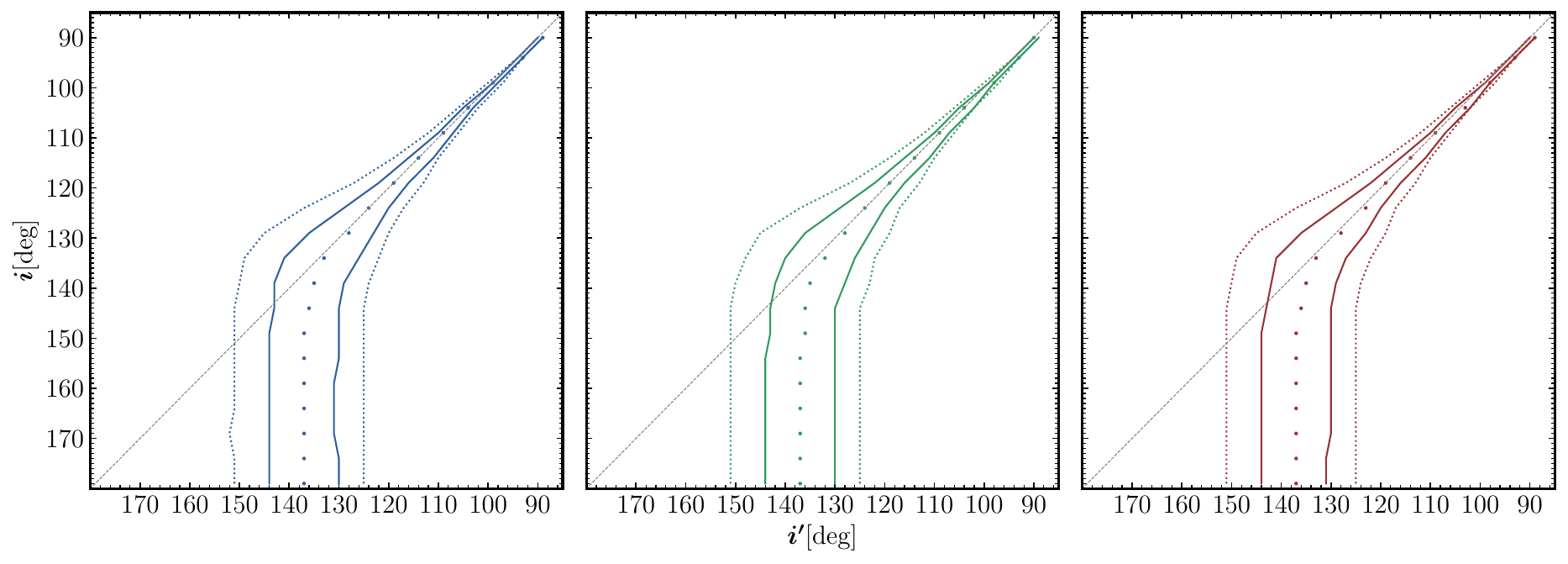}}\label{fig:ci2}
    {\includegraphics[width=1.0\textwidth]{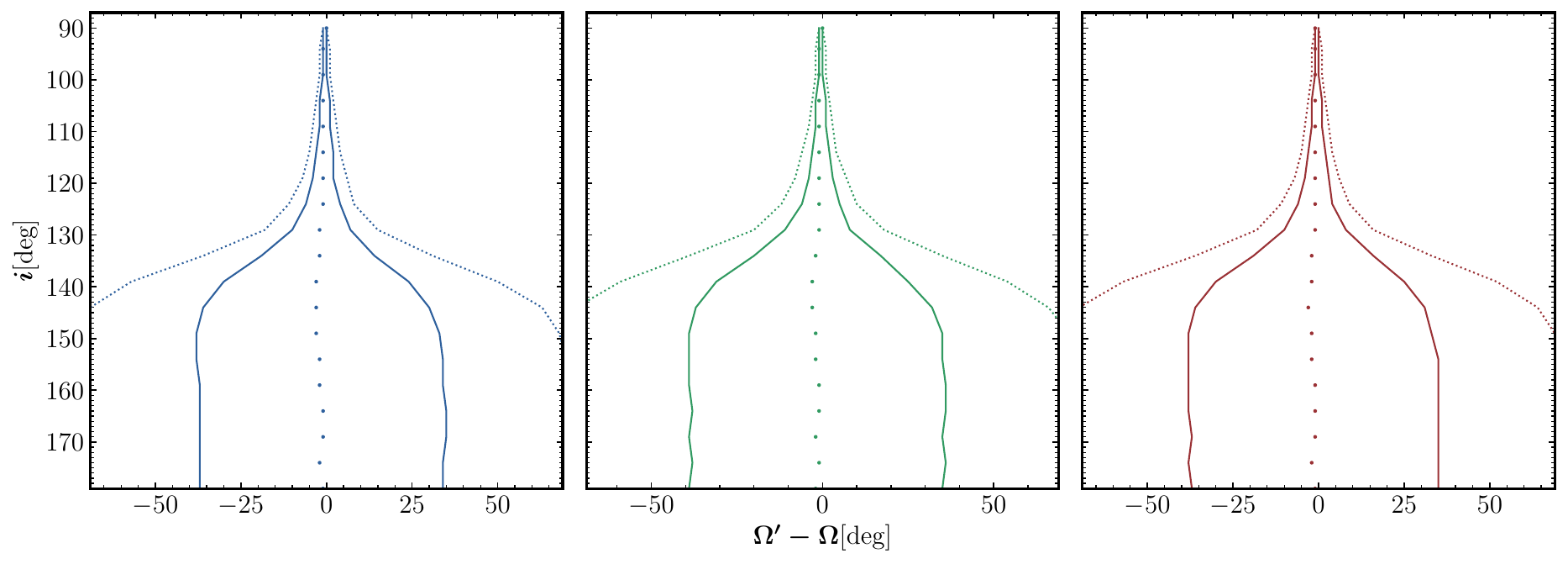}}\label{fig:co2} 
    \caption{Confidence intervals (solid lines for $\pm$1$\sigma$, and dotted lines for $\pm$2$\sigma$) of $i$ (upper panels), and $\Omega$ (lower panels) for $\gamma$ values of 316, 265, and 222 for the $B$, $V$, and $R$ passbands (left, middle, and right panels respectively). The dotted lines in upper panels illustrate critical values of $i$. Note, that the inclination range of $180^{\circ} - 90^{\circ}$ shown on the plots is fully equivalent to $0^{\circ} - 90^{\circ}$. If the orbital rotation as seen on the plane of the sky is clockwise, the BME method yields inclination value between $90^{\circ}$ and $180^{\circ}$.}
    \label{fig:ciio}
    \end{figure*}

    The presence of significant scatter in the polarization properties of stars within the NGC 6530 suggests that the region is dynamically complex, with varying distribution of interstellar dust and directions of interstellar magnetic field. This behavior is typical for a young open clusters, where stars are still forming and interacting with their surroundings. As a result, one can expect rather complex and strongly not uniform distribution of dust on the different lines of sight withing the cluster. For stars 7, 20, 21, and 22, the minimum degree of polarization is in the $V$ passband which is not typical for the IS polarization produced by homogeneous diffuse interstellar dust. We can suggest that in addition to variations in dust particles orientation and density, the variations in particles size and composition may also occur in the NGS 6530. We cannot also exclude that some of the field stars are intrinsically polarized. The most suspicious is the star 23 that shows sharp increase of polarization toward near infrared, e.g. from nearly zero value in the B-band to more than $1.0\%$ in the R-band.

    Although we were not able to estimate the value of IS polarization component for the HD 165052, we hope that our polarimetric observations of field stars provide insights into the structure and properties of the interstellar medium in the NGS 6530. The polarization data collected from field stars near HD 165052 can be used in future studies for examinations and modeling of the dynamic interaction between the interstellar dust, magnetic fields, and stellar community in the NGC 6530. 
    
    \subsection{Periodic Polarization Variability and Orbital Parameters}\label{sec:fcm}
 
    We used standard analytical method based on two-harmonics Fourier fit (commonly known as BME approach \citep{1978A&A....68..415B}) to fit the phase-folded curves of the Stokes $q$ and $u$ parameters. This method assumes a circular orbit with co-rotating light scattering envelope and the fit includes zeroth, first and second harmonics terms:

    \begin{equation}
    \begin{split}
    q =  q_0 + q_1 \cos \lambda + q_2 \sin \lambda + q_3 \cos 2\lambda + q_4 \sin 2\lambda, \\
    u =  u_0 + u_1 \cos \lambda + q_2 \sin \lambda + q_3 \cos 2\lambda + q_4 \sin 2\lambda,
    \end{split}
    \label{eq:fs}
    \end{equation}
	
    where $\lambda = 2 \pi \phi$, and $\phi$ represents the phase of the orbital period. The polarimetric data of Stokes $q$ and $u$ were phase-folded using the orbital period $P_{\text{orb}}$ [d] $= 2.95510$ and $T_0$ [HJD] $= 2449871.75$ from \citet{2002MNRAS.333..202A}. This was an obvious choice, because the value of $P_{\text{orb}}$ obtained by \citet{2002MNRAS.333..202A} is very close to that derived by us. Folding polarization data with the new value $P_{\text{orb}}$= 2.95585d obtained by \citep{2023MNRAS.521.2988R} revealed that it fails to represent adequately periodic variability of our polarization data set.  

    Bayesian inference method was employed to fit the model to the data points \citep{2018ApJS..236...11H}. Using this method on Stokes $q$ and $u$ data for the  $B$, $V$, and $R$ passbands, we determined the optimal Fourier coefficients. Subsequently, we performed curve fitting of the observed polarization data, as shown in Figure \ref{fig:fs}. The fit reveals an amplitude of variability of $\sim$0.10\%,  accompanied by some non-periodic scatter. We used \texttt{curve\_fit} function of the \texttt{scipy.optimize}\footnote{\url{https://docs.scipy.org/doc/scipy/reference/optimize.html}} library in \textsc{Python} to determine values of Fourier coefficients that were used as initial guess (priors) for our Bayesian interface, which enhanced the precision and subsequent values are given in Table \ref{table:fc}.     
                
    The orbital inclination $i$ and orbit's orientation in the sky, represented by the longitude of the ascending node $\Omega$ can be derived from the first ($q_{1,2}$, $u_{1,2}$) and second ($q_{3,4}$, $u_{3,4}$) harmonic terms of the Fourier series. Complete set of equations for computing orbital and other parameters are presented by \citet{1986ApJ...304..188D}. For additional details, please refer to Appendix \ref{sec:drissen}. However, in cases where the orbit is circular or nearly circular, and the distribution of scattering material is symmetric with respect to the orbital plane, the influence of the first harmonic terms becomes negligible. As is seen from Table \ref{table:fc}, this a clear case for HD 165052. Small orbit eccentricity has a very little / no effect on the fits, which are dominated by second harmonics. Thus, we have used only the second harmonic terms for deriving orbital parameters from the Fourier coefficients obtained for each passband.

    \begin{table}
    \centering
    \caption{Orbital parameters for HD~165052 derived from the polarization measurements in the $B$, $V$, and $R$ passbands. Parameters $A_{\rm q}$ and $A_{\rm u}$ are computed with Eq. \ref{Ratios}.} 
    \label{table:orbpar}
    \begin{tabular}[c]{ l c c c} 
    \hline\hline 
    Filter & $i$\tablefootmark{~(a, b)} & $\Omega$\tablefootmark{~(b)} & $A_{\rm q}/A_{\rm u}$ \\ \hline
    $B$ & $125^{\circ}$--$5^{\circ}$/+$55^{\circ}$ & $155^{\circ}+34^{\circ}/$--$37^{\circ}$ & 21.7/20.3 \\  
    & (55$^{\circ}$+5$^{\circ}$/--$55^{\circ}$) & ($335^{\circ}+34^{\circ}/$--$37^{\circ}$) & \\
    $V$ & 126$^{\circ}$--$5^{\circ}/+54^{\circ}$ & 133$^{\circ}+35^{\circ}/$--$38^{\circ}$/ & 18.3/10.6 \\  
    & (54$^{\circ}$+5$^{\circ}$/--$54^{\circ}$) & $(313^{\circ}+35^{\circ}/$--$38^{\circ})$ & \\
    $R$ & $123^{\circ}$--5$^{\circ}$/+$57^{\circ}$ & $146^{\circ}+35^{\circ}/$--$37^{\circ}$ & 17.9/21.8 \\  
    & (57$^{\circ}$+5$^{\circ}$/--$57^{\circ}$) & $(336^{\circ}+35^{\circ}/$--$37^{\circ})$ & \\
    \hline
    \end{tabular}\\
    \tablefoot{
    \tablefoottext{a}{Theses are de-biased values. The derived values through optimal Fourier coefficients, are 124$^{\circ}$(56$^{\circ}$), 125$^{\circ}$(55$^{\circ}$), and 122$^{\circ}$(58$^{\circ}$) in the $B$, $V$, and $R$ passbands, respectively.}
    \tablefoottext{b}{The given confidence intervals correspond to $\pm\sigma$.}}
    \end{table}

    The parameters $A_{\rm q}$ and $A_{\rm u}$ are the amplitude ratios between the second and first harmonics of Stokes $q$ and $u$, respectively. These parameters are quantifying the degree of a symmetry and concentration of scattering material around the orbital plane for a circular orbit. Expressions for computing these parameters are given in Appendix \ref{sec:drissen}.
    
    By applying Equations (\ref{Inclination} -- \ref{alphabets2}) from Appendix \ref{sec:drissen}, we computed the values of $i$, $\Omega$, $A_{\rm q}$, and $A_{\rm u}$ for the $B$, $V$, and $R$ passbands. Results are shown in Table \ref{table:orbpar}. The large values of $A_{\rm q}$ and $A_{\rm u}$ for all passbands indicate high degree of symmetry and/or concentration of light scattering material toward orbital plane in HD 165052.   

    \begin{figure}
    \centering
    \includegraphics[width=0.5\textwidth]{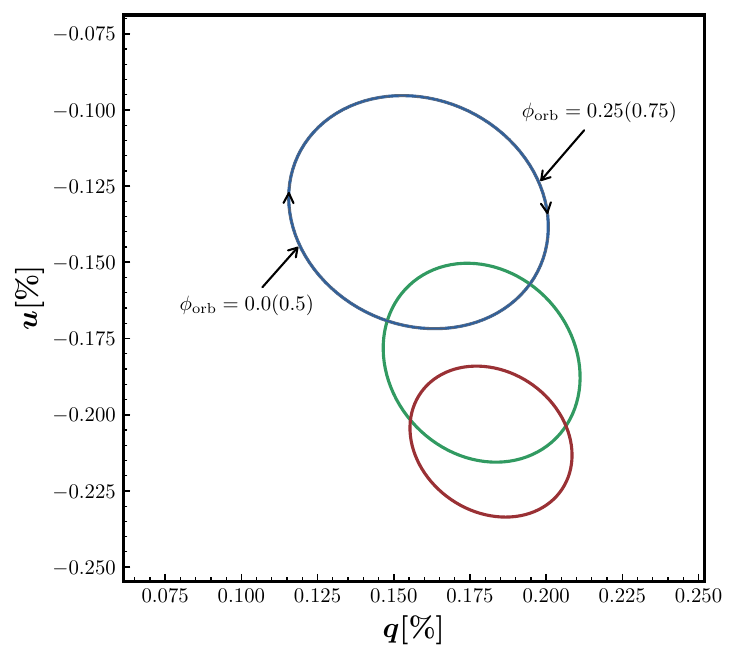}
    \caption{The variability of observed polarization for HD~165052 plotted on the Stokes ($q$, $u$) plane, represented by the ellipses of second harmonics of Fourier fit. Blue, green, and red colors represent $B$, $V$, and $R$ passbands respectively. The clockwise direction and phases of the orbital period are shown for the $B$ band ellipse. The angle between the major axis and the $q$-axis gives the orientation $\Omega$.} 
    \label{fig:elps}
    \end{figure}
    
    \begin{figure*}[!htp]
    \centering
    \includegraphics[width=1\textwidth]{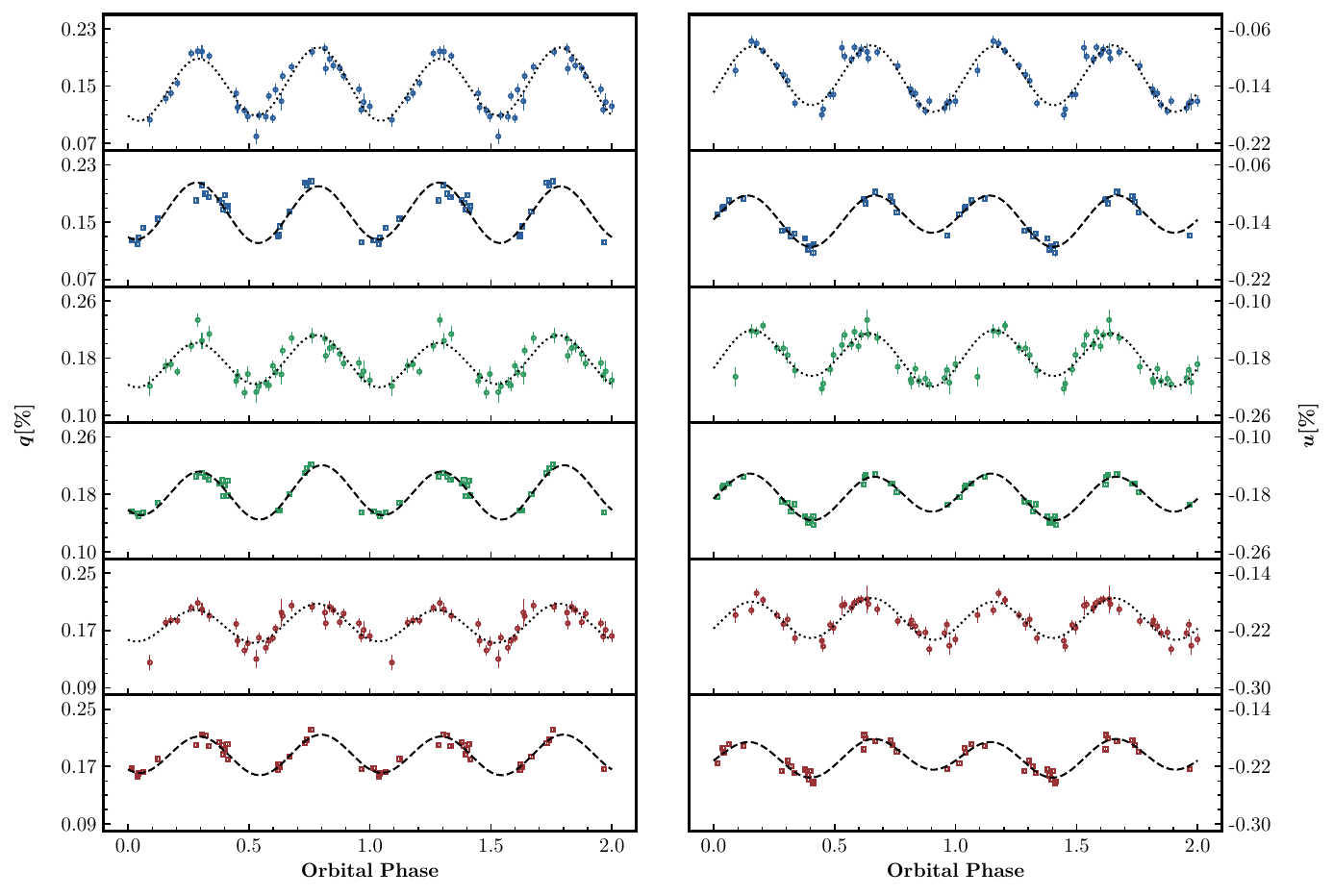}
    \caption{The variability of observed Stokes parameters $q$ and $u$ for HD~165052 in the $B$, $V$, and $R$ passbands (top two, middle two, and bottom two panels respectively), phase-folded with the orbital period of 2.95510d. The dotted Fourier fit curves and data points represented with circles show the data obtained in 2012; dashed Fourier fit curves and data points represented with squares show the data obtained in 2023. Best fit Fourier coefficients for two datasets are given in Table \ref{table:fcxy}. The lengths of the bars correspond to $\pm\sigma$ errors.} 
    \label{fig:fsxy}
    \end{figure*}

    \subsection{Confidence Intervals for the Orbital Parameters}
        
    As is evident from Figure \ref{fig:fs}, there is a noticeable non-periodic scatter around the fitting curves in all passbands. Moreover, because HD 165052 is a non-eclipsing system, the true inclination should be small. Thus, we should expect pronounced bias, asymmetric confidence intervals for the derived inclination $i$, and broad confidence intervals for $\Omega$.

    To determine the bias and corresponding confidence intervals for $i$ and $\Omega$ we have used a "figure of merit" parameter $\gamma$, introduced and defined by \citet{1994MNRAS.267....5W} as:
    
    \begin{equation}
    \gamma = \left(\frac{A}{\sigma_{\rm p}}\right)^2\frac{N}{2},
    \label{gamma}
    \end{equation}
    
    where, $A$ is the periodic fraction of the amplitude of polarization variability measured from the best fit and defined as:	
    \begin{equation}
    A = \frac{|q_{\rm max} - q_{\rm min}| + |u_{\rm max} - u_{\rm min}|}{4},
    \label{amplitude}
    \end{equation}
    
    $\sigma_{\rm p}$ is a standard deviation that is determined from the scatter of the observed Stokes parameters around the best fit curves, $N$ is the number of observations, and $q_{\rm max}$, $q_{\rm min}$, $u_{\rm max}$, $u_{\rm min}$ are maximum and minimum values of the fitted Stokes parameters $q$ and $u$. With our values of $\sigma_{\rm p}=$ 0.013, 0.011, 0.010 we have derived the corresponding values of $\gamma=$ 316, 265, 222 for the $B$, $V$, and $R$ passbands respectively.

    \begin{figure*}
    \centering
    {\includegraphics[width=1\textwidth]{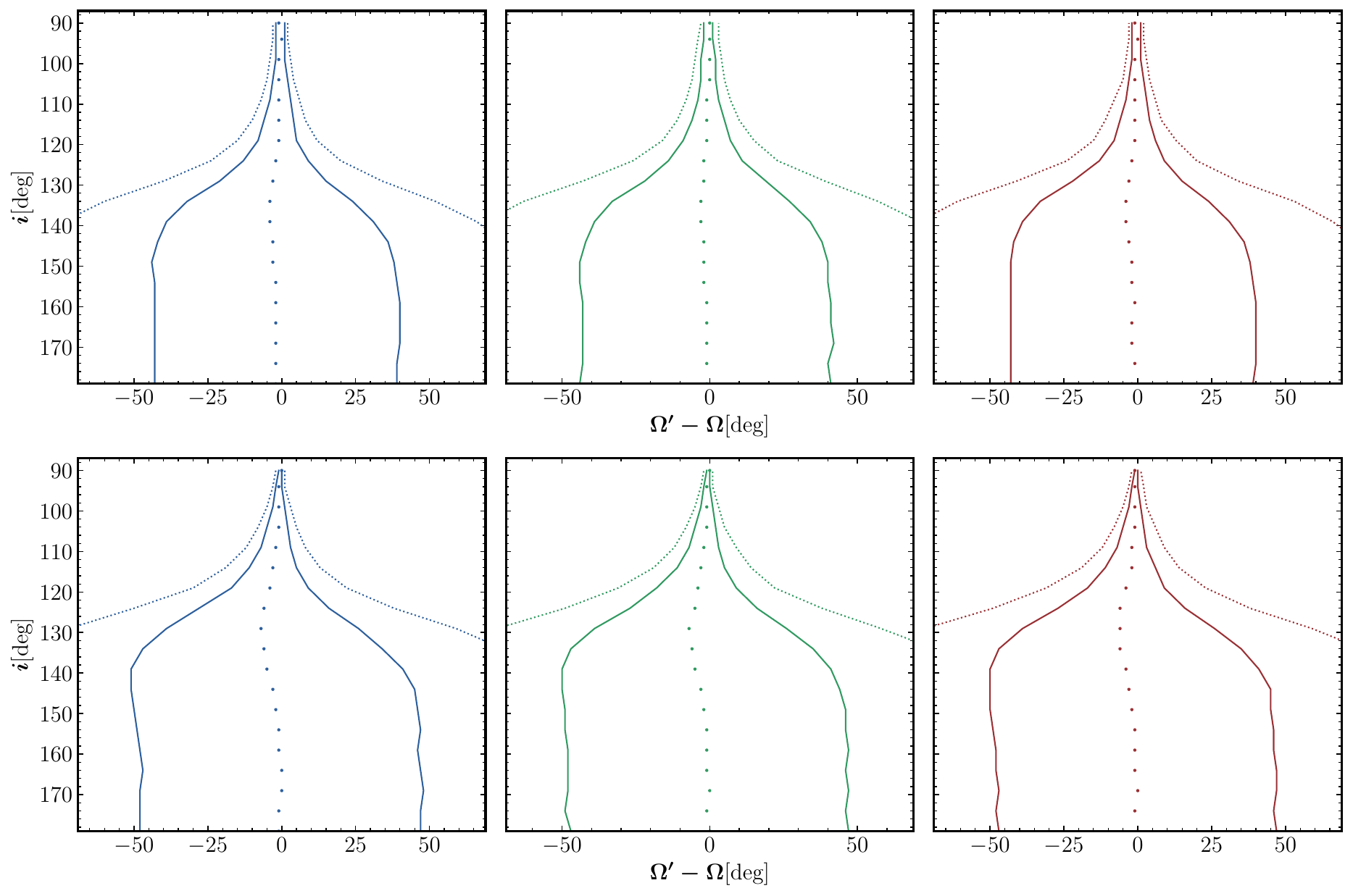}\label{fig:cioxy}} 
    \caption{Confidence intervals (solid lines for $\pm$1$\sigma$, and dotted lines for $\pm$2$\sigma$) of $\Omega$ for 2012 dataset (upper panels) with $\gamma$ values of 185, 70, and 72 for the $B$, $V$, and $R$ passbands (left, middle, and and right panels respectively), and for 2023 dataset (lower panels) with $\gamma$ values of 145, 240, and 171 for the $B$, $V$, and $R$ passbands (left, middle, and right panels respectively). Note, that confidence intervals for season 2012 are somewhat narrower due to the larger number of data points and, therefore, better phase coverage.}
    \label{fig:cioo2}
    \end{figure*}

    \begin{table*}[!htb]
    \centering
    \caption{Best fit Fourier coefficients for Stokes $q$ and $u$ for data obtained in 2012 (upper panel) and 2023 (lower panel).}
    \label{table:fcxy}
    \centering
    \scalebox{1.0}{
    \begin{tabular}[c]{ l c c c c c c c c c c }
    \hline\hline 
    \tabhead{\shortstack{\rm Filter }} & \tabhead{\shortstack{\rm $q_0$}}  & \tabhead{\shortstack{\rm $q_1$}} 
    & \tabhead{\shortstack{\rm $q_2$}} & \tabhead{\shortstack{\rm $q_3$}} & \tabhead{\shortstack{\rm $q_4$}}
    & \tabhead{\shortstack{\rm $u_0$}} & \tabhead{\shortstack{\rm $u_1$}} & \tabhead{\shortstack{\rm $u_2$}}
    & \tabhead{\shortstack{\rm $u_3$}} & \tabhead{\shortstack{\rm $u_4$}} \\

    \hline
    $B$ & 0.1506 & -0.0016 & -0.0084 & -0.0402 & -0.0210
    & -0.1277 & -0.0043 & 0.0019 & -0.0169 & 0.0403 \\
    & $\pm$0.0024 & $\pm$0.0030 & $\pm$0.0035 & $\pm$0.0035 & $\pm$0.0032 
    & $\pm$0.0019 & $\pm$0.0030 & $\pm$0.0027 & $\pm$0.0028 & $\pm$0.0028 \\	
				
    $V$ & 0.1743 & -0.0011 & -0.0057 & -0.0300 & -0.0124
    & -0.1777 & -0.0047 & 0.0062 & -0.0116 & 0.0325 \\
    & $\pm$0.0025 & $\pm$0.0035 & $\pm$0.0040 & $\pm$0.0039 & $\pm$0.0034 
    & $\pm$0.0022 & $\pm$0.0032 & $\pm$0.0027 & $\pm$0.0031 & $\pm$0.0033 \\	

    $R$ & 0.1783 & 0.0017 & -0.0041 & -0.0230 & -0.0087
    & -0.2046 & -0.0025 & -0.0013 & -0.0103 & 0.0252 \\
    & $\pm$0.0018 & $\pm$0.0026 & $\pm$0.0023 & $\pm$0.0025 & $\pm$0.0025 
    & $\pm$0.0016 & $\pm$0.0024 & $\pm$0.0025 & $\pm$0.0022 & $\pm$0.0023 \\				
    
    \hline	

    $B$ & 0.1630 & 0.0018 & 0.0030 & -0.0357 & -0.0172
    & -0.1339 & 0.0080 & -0.0058 & -0.0106 & 0.0289 \\
    & $\pm$0.0025 & $\pm$0.0040 & $\pm$0.0049 & $\pm$0.0049 & $\pm$0.0045
    & $\pm$0.0013 & $\pm$0.0022 & $\pm$0.0023 & $\pm$0.0021 & $\pm$0.0021 \\
				
    $V$ & 0.1820 & 0.0042 & -0.0033 & -0.0280 & -0.0196
    & -0.1818 & 0.0063 & -0.0015 & -0.0106 & 0.0264 \\
    & $\pm$0.0017 & $\pm$0.0028 & $\pm$0.0030 & $\pm$0.0028 & $\pm$0.0029 
    & $\pm$0.0010 & $\pm$0.0017 & $\pm$0.0018 & $\pm$0.0016 & $\pm$0.0019 \\	

    $R$ & 0.1863 & 0.0019 & -0.0008 & -0.0224 & -0.0148
    & -0.2069 & 0.0033 & -0.0049 & -0.0082 & 0.0216 \\
    & $\pm$0.0016 & $\pm$0.0026 & $\pm$0.0029 & $\pm$0.0027 & $\pm$0.0027 
    & $\pm$0.0017 & $\pm$0.0027 & $\pm$0.0030 & $\pm$0.0028 & $\pm$0.0026 \\				
    \hline	
    \end{tabular}}
    \end{table*}

    Due to the unavoidable presence of noise in real polarimetric data, the inclination angle \(i\) derived from Fourier fits is consistently biased towards higher values (\citet{1981MNRAS.194..283A}; \citet{1982MNRAS.198...45S}; \citet{1994MNRAS.267....5W}). This bias is particularly pronounced for lower true values of \(i\), making it increasingly challenging to constrain smaller inclinations. Real non-periodic fluctuations present in polarization variability will also introduce a similar bias, as noted by \citet{2000AJ....120..413M}.  

    A critical limitation arises when the true inclination is low (\(i < 45^\circ\)), especially for low-eccentricity orbits, as demonstrated by \citet{2000AJ....120..413M}. In such cases, the BME method cannot reliably recover the lower inclination values, resulting in systematically overestimated inclinations. Importantly, this bias also skews the confidence intervals for \(i\), leading to asymmetric \(1\sigma\) and \(2\sigma\) intervals. The lower boundary of these intervals extends disproportionately towards smaller \(i\), often reaching \(i = 0^\circ\) for systems with low inclinations. For non-eclipsing systems such as HD 165052, this means that polarimetric data can provide only an upper limit for the true inclination, effectively excluding the possibility of constraining smaller values.
    
    Moreover, as the true inclination decreases, the confidence intervals for the position angle of the ascending node (\(\Omega\)) expand significantly \citep{1994MNRAS.267....5W}. This systematic bias and its impact highlight the inherent difficulty in deducing orbital parameters for low inclination orbits from polarimetric data using Fourier fits.

    \begin{table}
    \centering
    \caption{Orbital parameter $\Omega$ for HD~165052 derived for thee $B$, $V$, and $R$ passbands. The upper panel values are obtained for dataset from year 2012 and the lower panel values are from year 2023. }
    \label{table:orbparxy}
    \begin{tabular}[c]{ l c } 
    \hline\hline 
    Filter & $\Omega$\tablefootmark{~(a)} \\ \hline
    $B$ & $328^{\circ}(148^{\circ})+39^{\circ}/-43^{\circ}$ \\  
    $V$ & $283^{\circ}(103^{\circ})+41^{\circ}/-44^{\circ}$ \\  
    $R$ & $258^{\circ}(78^{\circ})+40^{\circ}/-43^{\circ}$ \\  
    \hline
    $B$ & $170^{\circ}(350^{\circ})+47^{\circ}/-48^{\circ}$ \\  
    $V$ & $154^{\circ}(334^{\circ})+47^{\circ}/-49^{\circ}$ \\  
    $R$ & $152^{\circ}(332^{\circ})+47^{\circ}/-47^{\circ}$ \\  
    \hline
    \end{tabular}\\
    \tablefoot{
    \tablefoottext{a}{The given confidence intervals correspond to $\pm\sigma$.}}
    \end{table}

    Using our values of \(\gamma\) and \(N\), the number of data points with their corresponding orbital phases and measurement errors, we conducted our own Monte Carlo simulations to determine the \(1\sigma\) and \(2\sigma\) confidence intervals for \(i\) and \(\Omega\) (see Figure \ref{fig:ciio}). This methodology was adapted from \citet{1994MNRAS.267....5W}, but instead of relying solely on simulated data as done by \citet{1994MNRAS.267....5W} and \citet{2000AJ....120..413M}, we incorporated the actual errors and phase sampling from our data set. This approach ensures that our results are directly tied to the specific characteristics of our observations, rather than generalized assumptions about noise or measurement uncertainties.

    In our simulations, we modeled Stokes parameters for values of \(i\) ranging from \(180^\circ\) to \(90^\circ\) with a decreasing step of \(5^\circ\), using the standard \citet{1978A&A....68..415B} model (Eq. \ref{eq:fs}). This inclination range is fully equivalent to the "normal" range of \(0^\circ\)–\(90^\circ\) and reflects the fact that the orbital rotation in the HD 165052 binary system is clockwise. Gaussian noise was then simulated in the Stokes parameters \(q\) and \(u\) by adding fluctuations with variance \(\sigma^2 = 0.5 N A^2 / \gamma\). The Fourier model (Eq. \ref{eq:fs}) was subsequently fit to the simulated data using the Bayesian approach identical to the one used for deriving the Fourier coefficients listed in Table \ref{table:fc}.
    
    From the simulated Fourier coefficients, we derived \(i^\prime\) using Eq. \ref{Inclination}. Similarly, for confidence intervals of \(\Omega\), we determined \(\Omega^\prime - \Omega\), where \(\Omega^\prime\) was deduced using Eq. \ref{Orientation}, with the input value \(\Omega = 0\) chosen for simplicity. For each input value of \(i\), we repeated the simulations to obtain 100 values of \(i^\prime\) and \(\Omega^\prime - \Omega\). This iterative approach allowed us to robustly quantify the confidence intervals and biases associated with our data, providing realistic constraints on the orbital parameters derived from our polarimetric measurements.

    Our confidence intervals, shown in Figure \ref{fig:ciio}, are noticeably narrower than those presented by \citet{1994MNRAS.267....5W} for similar values of \(\gamma\). This difference is primarily due to the larger number of observations in our dataset (\(N = 68\)) compared to the \(N = 23\) used by \citet{1994MNRAS.267....5W}. A larger \(N\) provides better phase coverage, resulting in narrower confidence intervals. While \(\gamma\) quantifies the precision of the Fourier fit, the density of phase sampling also plays a critical role in determining the width of the confidence intervals. Consequently, increasing \(N\) enhances phase sampling density, reducing the confidence interval width even for smaller values of \(\gamma\). The slight asymmetry observed in our confidence intervals can be attributed to the uneven phase sampling inherent in real observations, unlike the evenly distributed phase coverage employed by \citet{1994MNRAS.267....5W}.
    
    Our derived inclinations for all passbands (see Table \ref{table:orbpar}) are close to the critical value of \(i\) shown in Figure \ref{fig:ciio}. Beyond this threshold, the lower limit of the \(1\sigma\) confidence interval extends to \(i = 180^{\circ}\) (or \(i = 0^{\circ}\)); see \citet{1994MNRAS.267....5W}. This indicates that our polarization data analysis allows us to establish only an upper limit of \(i \simeq 55^{\circ}\) for the orbital inclination of HD 165052.
    
    To determine the confidence intervals for \(\Omega\), we set the true inclination angle to \(i = 22.7^{\circ}\), as determined by \citet{2023MNRAS.521.2988R}. This estimate, based on spectroscopic analysis and atmospheric modeling, is certainly more realistic than that derived from polarimetric data. O-type stars typically have masses exceeding \(20\,M_{\odot}\). The values of \(M \sin^3 i\) derived from spectroscopy by \citet{2023MNRAS.521.2988R} and others, ranging from \(1.24\,M_{\odot}\) to \(1.37\,M_{\odot}\), imply that inclinations greater than \(23.1^{\circ}\) would result in masses below the \(20\,M_{\odot}\) threshold. The confidence intervals for \(\Omega\), derived using this inclination, are presented in Table \ref{table:orbpar}. Despite large uncertainties, the values of \(\Omega\) across different wavelengths are consistent, yielding an average orbital orientation angle on the plane of the sky of approximately \(145^{\circ}\) (\(\sim 325^{\circ}\)). Note that \(\Omega\) determined from polarization data inherently carries a \(\pm180^{\circ}\) ambiguity, as reflected in Table \ref{table:orbpar}.
    
    Figure \ref{fig:elps} illustrates the ellipses of the second harmonics on the \((q, u)\) plane for the \(B\), \(V\), and \(R\) passbands. The degree of eccentricity of these ellipses is directly related to the orbital inclination \(i\), while the orientation of their major axes with respect to the \(q\)-axis corresponds to the orbital orientation \(\Omega\) on the sky plane. The direction of traversal along the ellipses indicates the direction of orbital motion on the celestial plane. Given the derived inclination range of \(90^\circ \leq i \leq 180^\circ\), the binary system's rotation on the plane of the sky is clockwise.

    \subsection{Effect of the Apsidal Motion}
    
    HD 165052 has been known to exhibit apsidal precession (\citet{2002MNRAS.333..202A}; \citet{2013MNRAS.433.1300F}; \citet{2023MNRAS.521.2988R}) that can be utilized for confirming the predictions of general relativity theory, as the system's high mass and proximity make relativistic effects more pronounced. \citet{2013MNRAS.433.1300F} determined the apsidal motion rate of the system to be 12.1 $\pm$ 0.3$^{\circ}$yr$^{-1}$ and recently \citet{2023MNRAS.521.2988R} derived the value of 11.3 + 0.64 / - 0.49$^{\circ}$yr$^{-1}$. 

    We benefit from our extensive polarimetric observational history of HD 165052, spanning nearly 11 years. Our dataset comprises 33 data points acquired from the KVA telescope in 2012 and an additional 23 data points collected from the T60 telescope in 2023 (Table \ref{table:log}). To investigate possible effect of apsidal motion through polarimetry, we conducted separate Fourier fits on these two datasets. Best fits are shown on Figure \ref{fig:fsxy} and the corresponding Fourier coefficients are presented in Table \ref{table:fcxy}. As is seen from Table \ref{table:fcxy}, the effect of apsidal motion on the polarization variability, if present, appears to be very subtle. Only for Stokes $u$ in the B-band the difference in Fourier term $u_{4}$ between two datasets is larger than $3\sigma$. The presence of non-periodic noise in both fits is also obvious. 
    
    Nevertheless, we have tried to derive the separate values of $\Omega$ for two sets of polarization data in order to estimate the significance of the difference between them. For estimation of confidence intervals for the values of $\Omega$ we have employed the method described in the previous section. The corresponding $1\sigma$ and $2\sigma$ confidence intervals are shown on Figure \ref{fig:cioo2}. The resulting values of $\Omega$ are presented in Table \ref{table:orbparxy}. As is seen from Figure \ref{fig:cioo2}, $2\sigma$ confidence intervals for derived $\Omega$ for the orbit inclination $i$ above $135^{\circ}$ (or below $45^{\circ}$) are extending up to ($-180^{\circ}, +180^{\circ}$). Thus, although analysis of our polarization data do show some evidence for the apsidal motion, we cannot reliably estimate its rate due to large uncertainty in value of $\Omega$ derived from our polarization data.
    
    \subsection{Determination of the mass-loss rate}

    Because HD 165052 is an O-type binary, the most likely cause for the polarization variations is the electron scattering in the stellar winds of the hot components. The strong evidence for presence of the hot shocked gas region due to wind-to-wind interaction in HD 165052 was already presented by \citet{2002MNRAS.333..202A} from their analysis of ROSAT data. The presence of a real stochastic variability in polarization of HD 165052 is also quite obvious. The natural explanation for this phenomenon is highly clumped radiatively driven stellar winds which is not uncommon in early-type binaries consisting of O-type components. 
    
    It is possible to estimate the mass-loss rate from the component(s) in an early-type binary using polarization variability amplitude, employing the method proposed by \citet{1988ApJ...330..286S}. The method utilizes the polarization variability amplitude \(A_{\rm p}\), defined by \citet{1978A&A....68..415B} as:

    \begin{equation}
        A_{\rm p} = \tau_{0} H (1 + \cos^2 i),
        \label{ap_original}
    \end{equation}
    
    where \(\tau_{0} H\) is a measure of the optical depth moment and the scattering geometry of the stellar wind, and \(i\) is the orbital inclination. The values of \(A_{\rm p}\) determined from our Fourier fits are 0.059\%, 0.046\%, and 0.039\% for the \(B\), \(V\), and \(R\) passbands, respectively. Although electron scattering is independent of wavelength and should yield equal amplitudes, differences may arise from dilution effects caused by unpolarized radiation from the gaseous disk/envelope. 
    
    As shown by \citet{1988ApJ...330..286S}, in the case or Wolf-Rayet (WR)+O system, the mass-loss rate due to stellar wind from the primary component can be estimated using the following formula:
    
    \begin{equation}
        \dot{M} = \frac{(16\pi)^2 m_{\rm p} v_{\infty} a A_{\rm p}}{(1 + \cos^2 i) 3 \sigma_{\rm t} f_{\rm c} I},
        \label{mass-loss}     
    \end{equation}
    
    where \(\dot{M}\) is the mass-loss rate, \(f_{\rm c}\) is the fraction of total light from the companion star, \(v_{\infty}\) is the wind terminal velocity, \(a\) is the semi-major axis, \(m_{\rm p}\) is mass of proton, \(\sigma_{\rm t}\) is Thomson cross section, and \(I\) is a specific integral. Evaluating this integral requires choosing a specific wind velocity law characterized by the parameter \(\beta\). In mass-loss rate calculations for massive stars in binary systems, the polarization amplitude \( A_{\rm p} \) is directly related to the wind properties and mass-loss rates of the stars. The observed polarization in a binary system is influenced by the combined light from both stars, but the manner in which each star contributes depends on the system type.

    In a WR+O binary system, the WR star's strong stellar wind serves as the primary source of polarization, while the O-type companion, despite contributing to the total light of teh binary system, has a much weaker wind. Consequently, the observed polarization is primarily due to photons from the O-star scattering off free electrons in the WR wind. The WR star’s light originates from a dense, optically thick wind, where multiple scattering events occur, effectively averaging out any intrinsic polarization under the assumption of spherical symmetry. In contrast, the companion’s light travels asymmetrically through the WR wind, undergoing single scattering, which generates net polarization. However, the observed polarization is diluted by the unpolarized light from the WR star. To correct for this dilution when deriving the WR star’s mass-loss rate (\(\dot{M}\)), the observed polarization amplitude (\(A_{\rm p}\)) is divided by the fractional light contribution of the O-star (\(f_{\rm c}\)).
    
    In contrast, for an O+O binary system, both stars contribute comparably to the observed polarization. Since both stars have similar luminosity and wind strength, their individual contributions to the total polarization and dilution are roughly equal. Therefore, we assume \( f_{\rm c} = 0.5 \). Moreover, we also account for the difference in wind composition between the O-type star wind and the WR star wind. The wind of the WR star is composed of He, while the O-type star wind is composed of both H and He, but is mostly dominated by H. In the case of fully ionized He, each He nucleus contains 2 electrons per 4 nucleons, i.e., \( \alpha_{\rm WR} = 2/4 = 0.5 \) as given by \citet{1988ApJ...330..286S}. For fully ionized H, the ratio is 1:1. We assume that the O-star wind consists of 90\% of H and 10\% of He, and we have electron / nucleon ratio $\sim$$1.0$. Thus, the formula for the mass-loss rate per component in HD 165052 system becomes:
    
    \begin{equation}
    \begin{split}
        \dot{M} &= \frac{(16\pi)^2 m_{\rm p} v_{\infty} a A_{\rm p} }{(1 + \cos^2 i) 3 \sigma_{\rm t} f_{\rm c}I} \cdot \frac{1}{2}, \\
        \text{or} \\
        \dot{M} \, [M_{\odot} \; {\rm yr^{-1}}] &= \frac{1.16 \times 10^{-7} v_{\infty} \, ({\rm km \; s^{-1}}) \, a \, (R_{\odot}) \, A_{\rm p} \,}{(1 + \cos^2 i) \, f_{\rm c}I}.
    \end{split}
    \label{mass-loss-o}
    \end{equation}
        
    We can now estimate \(\dot{M}\) by using the observed polarization variability amplitude \(A_{\rm p}\) and the orbital inclination \(i = 22.7^{\circ} \pm 0.6^{\circ} \) from \citet{2023MNRAS.521.2988R}. The wind velocity \(v_{\infty} = 2335 \, \text{km} \, \text{s}^{-1}\) for both components, as determined from combined IUE spectra by \citet{1997MNRAS.284..265H}, is used. 
    
    For primary component: \(a_{\rm p} \sin i = 5.66 \pm 0.04~R_{\odot}\), \(R_{*,\rm p} = 7.0 + 0.5 / - 0.4~R_{\odot}\), give \(a_{\rm p}/R_{*,\rm p} = 2.1\), and for secondary component: \(a_{\rm s} \sin i = 6.27 \pm 0.06~R_{\odot}\), \(R_{*,\rm s} = 6.2 + 0.4 / - 0.3~R_{\odot}\), give \(a_{\rm s}/R_{*,\rm s} = 2.6\) \citep{2023MNRAS.521.2988R}. We assumed \(\epsilon = R_{\rm i}^{\prime}/R_{*} = 1.0\) (optically thin scattering envelope). We chose \(\beta = 0.8\), as suggested by \citet{1986ApJ...303..239A} and \citet{1991ApJ...366..308C} for O-type stars. By using these values, we selected integral values of \(I_{\rm p} = 12.6\) and \(I_{\rm s} = 12.3\) from the plot given by \citet{1988ApJ...330..286S} (Fig. 9 therein) for primary and secondary components respectively. In order to obtain uncertainty estimates on derived values of $\dot{M}$, we used the method of error propagations, as suggested by \citet{1988ApJ...330..286S}.

    For each observed band ($B$, $V$, $R$), we calculated the mass-loss rates for both stars and the total mass-loss rate is then obtained by summing these contributions. Uncertainties for each component are also propagated, providing a robust estimate of the total mass-loss rate for the system:

    \begin{itemize}
        \item \textbf{B-band:}
        \begin{itemize}
            \item Primary: \( 2.010 \times 10^{-7} \pm 2.214 \times 10^{-8}M_{\rm \odot} \; \rm yr^{-1} \),
            \item Secondary: \( 2.255 \times 10^{-7} \pm 2.496 \times 10^{-8}M_{\rm \odot} \; \rm yr^{-1} \),
            \item Total: \( 4.265 \times 10^{-7} \pm 3.337 \times 10^{-8}M_{\rm \odot} \; \rm yr^{-1} \).
        \end{itemize}
        \item \textbf{V-band:}
        \begin{itemize}
            \item Primary: \( 1.567 \times 10^{-7} \pm 1.725 \times 10^{-8}M_{\rm \odot} \; \rm yr^{-1} \),
            \item Secondary: \( 1.758 \times 10^{-7} \pm 1.945 \times 10^{-8}M_{\rm \odot} \; \rm yr^{-1} \),
            \item Total: \( 3.325 \times 10^{-7} \pm 2.600 \times 10^{-8}M_{\rm \odot} \; \rm yr^{-1} \).
        \end{itemize}
        \item \textbf{R-band:}
        \begin{itemize}
            \item Primary: \( 1.329 \times 10^{-7} \pm 1.463 \times 10^{-8}M_{\rm \odot} \; \rm yr^{-1} \),
            \item Secondary: \( 1.491 \times 10^{-7} \pm 1.649 \times 10^{-8}M_{\rm \odot} \; \rm yr^{-1} \),
            \item Total: \( 2.819 \times 10^{-7} \pm 2.204 \times 10^{-8}M_{\rm \odot} \; \rm yr^{-1} \).
        \end{itemize}
    \end{itemize}

    This gives the average value over three passbands as  \( 3.47 \times 10^{-7} \pm 1.59 \times 10^{-8}M_{\rm \odot} \; \rm yr^{-1} \). Our estimation of $\dot{M}$ is somewhat higher than previously derived mass-loss rate of $\dot{M} = 2.24 \times 10^{-7} M_{\rm \odot} \; \rm yr^{-1}$ published by \citet{1991ApJ...368..241C}. The decrease of $\dot{M}$ from the B to R passbands is due to decrease of \(A_{\rm p}\) with the wavelengths. Thomson scattering is a gray process, but there is an additional dilution of scattered polarized radiation due to free-free emission arising in the gaseous disk / shell. Contribution from this emission increases towards the near-infrared. Thus, we expect that our "upper" estimate obtained for the B-band, \( 4.26 \times 10^{-7} \pm 3.34 \times 10^{-8}M_{\rm \odot} \; \rm yr^{-1} \), is least affected by this dilution and probably is more reliable. 

    \section{Conclusions}\label{sec:conclusions}	
    Our comprehensive polarimetric study of the O+O spectral type binary HD 165052 has clearly revealed the presence of the periodic polarization variability. Lomb-Scargle periodograms unambiguously determined the orbital period value of 2.95510~d. This value is the good agreement with the previously determined values of the orbital period 2.95510~d \citet{2002MNRAS.333..202A} and 2.95515~d \citet{2007A&A...474..193L}. The most plausible mechanism of polarization variability is the Thomson scattering of light in the interacting stellar winds. Our polarimetry of the field stars in the near vicinity of HD 165052 revealed complex and strongly non-uniform pattern of interstellar polarization in the cluster NGC 6530. Our multi-wavelengths observations show that in addition to variations of dust particles density and orientations, variation in particles size and composition may occur. 
    
    As we have found, polarization variability in HD 165052 is dominated by the second harmonics of the orbital period. This indicates a nearly symmetric scattering material geometry and a high concentration of light-scattering material toward the orbital plane. Based on a refined Fourier fit, we derived the average values of the orbital parameters: orbital inclination \( i = 55^{\circ} + 5^{\circ} / -55^{\circ} \), and \(\Omega = 148^{\circ} (328^{\circ}) + 20^{\circ} / -22^{\circ} \), averaged over the \( B \), \( V \), and \( R \) passbands. The direction of orbital motion in the sky is clockwise. Using the values of polarization periodic variability amplitude, we estimated the total mass-loss rate of the binary system as \( \dot{M} = 3.47 \times 10^{-7} \pm 1.59 \times 10^{-8}M_{\rm \odot} \; \rm yr^{-1} \).

    We have demonstrated that BME analytic solution is not able to put a reliable constrain on inclination of the orbit for the low inclination systems in the presence of non-periodic noise in polarization variability. Nevertheless, high-precision polarization data can be still used for obtaining reliable independent estimates for the orbital period and mass loss rates. Numerical modeling may provide a better estimates, because it is accounting for component sizes, their separation, and relevant scattering scenarios.

    \begin{acknowledgements}
    This work was supported by the ERC Advanced Grant Hot-Mol ERC-2011-AdG-291659 (www.hotmol.eu). Dipol-2 was built in the cooperation between the University of Turku, Finland, and the Kiepenheuer Institutf\"{u}r Sonnenphysik, Germany, with the support by the Leibniz Association grant SAW-2011-KIS-7. We are grateful to the Institute for Astronomy, University of Hawaii for the observing time allocated for us on the T60 telescope at the Haleakal\={a} Observatory. All raw data and calibrations images are available on request from the authors.

    \end{acknowledgements}

    \bibliography{allbib}
    \bibliographystyle{aa}

    \appendix
    \section{Tables}\label{sec:logs}

    \begin{table}[htp!]
    \caption{Log of polarimetric observations for HD 165052.} 
    \label{table:log}
    \centering
    \begin{tabular}[c]{l  c  c  c c }
    \hline\hline 
    Date & MJD & $T_{\rm exp}[\rm s]$ &  $N_{\rm obs}$ & Telescope\\ \hline
    2012--07--28 & 56136.43 & 640 & 64 & KVA \\
    2012--07--30 & 56138.44 & 640 & 64 & KVA \\
    2012--07--31 & 56139.44 & 640 & 64 & KVA \\
    2012--08--01 & 56140.42 & 640 & 64 & KVA \\
    2012--08--06 & 56145.42 & 640 & 64 & KVA \\
    2012--08--07 & 56146.42 & 640 & 64 & KVA \\
    2012--08--08 & 56147.43 & 640 & 64 & KVA \\
    2012--08--27 & 56166.38 & 640 & 64 & KVA \\
    2012--08--28 & 56167.43 & 640 & 64 & KVA \\
    2012--08--29 & 56168.36 & 640 & 64 & KVA \\
    2012--09--03 & 56173.36 & 640 & 64 & KVA \\
    2012--09--04 & 56174.43 & 640 & 64 & KVA \\
    2012--09--06 & 56176.44 & 640 & 64 & KVA \\
    2012--09--09 & 56179.36 & 640 & 64 & KVA \\
    2012--09--10 & 56180.35 & 640 & 64 & KVA \\
    2012--09--11 & 56181.35 & 640 & 64 & KVA \\
    2012--09--13 & 56183.35 & 640 & 64 & KVA \\
    2012--09--14 & 56184.37 & 640 & 64 & KVA \\
    2012--09--16 & 56186.35 & 640 & 64 & KVA \\
    2012--09--18 & 56188.37 & 640 & 64 & KVA \\
    2012--09--20 & 56190.36 & 640 & 64 & KVA \\
    2012--09--21 & 56191.35 & 640 & 64 & KVA \\
    2012--09--22 & 56192.34 & 640 & 64 & KVA \\
    2012--09--25 & 56195.34 & 640 & 64 & KVA \\
    2012--09--27 & 56197.35 & 640 & 64 & KVA \\
    2012--09--30 & 56200.34 & 640 & 64 & KVA \\
    2012--10--02 & 56202.35 & 640 & 64 & KVA \\
    2012--10--03 & 56203.34 & 640 & 64 & KVA \\
    2012--10--06 & 56206.33 & 640 & 64 & KVA \\
    2012--10--08 & 56208.34 & 640 & 64 & KVA \\
    2012--10--10 & 56210.33 & 640 & 64 & KVA \\
    2012--10--12 & 56212.33 & 640 & 64 & KVA \\   
    2015--05--14 & 57156.09 & 640 & 64 & T60 \\
    2015--05--15 & 57157.06 & 640 & 64 & T60 \\
    2015--05--19 & 57161.04 & 640 & 64 & T60 \\
    2015--05--19 & 57161.99 & 640 & 56 & T60 \\
    2015--05--21 & 57163.98 & 640 & 64 & T60 \\
    2015--05--22 & 57164.98 & 640 & 64 & T60 \\
    2015--08--11 & 57245.76 & 480 & 48 & T60 \\
    2015--08--12 & 57246.74 & 640 & 64 & T60 \\
    2015--08--22 & 57256.80 & 640 & 64 & T60 \\
    2016--04--21 & 57499.01 & 480 & 48 & T60 \\
    2016--04--23 & 57501.03 & 640 & 64 & T60 \\
    2016--04--25 & 57503.03 & 640 & 64 & T60 \\
    2016--04--26 & 57504.01 & 640 & 64 & T60 \\
    2023--04--08 & 60042.08 & 640 & 64 & T60 \\
    2023--04--10 & 60044.04 & 640 & 64 & T60 \\
    2023--04--11 & 60045.04 & 640 & 64 & T60 \\
    2023--04--16 & 60050.06 & 640 & 64 & T60 \\
    2023--04--16 & 60050.97 & 640 & 64 & T60 \\
    2023--04--17 & 60051.97 & 640 & 64 & T60 \\
    2023--04--18 & 60052.97 & 640 & 64 & T60 \\
    2023--04--24 & 60058.03 & 640 & 64 & T60 \\
    2023--04--24 & 60058.96 & 800 & 80 & T60 \\
    \end{tabular}
    \end{table}

    \begin{table}[htp!]
    \centering
    \begin{tabular}[c]{l  c  c  c c }
    \\
    \\
    \\
    \\
    \\
    \\
    2023--04--25 & 60059.95 & 640 & 64 & T60 \\
    2023--05--01 & 60065.04 & 640 & 64 & T60 \\
    2023--05--02 & 60066.96 & 640 & 64 & T60 \\
    2023--05--03 & 60067.96 & 640 & 64 & T60 \\
    2023--05--05 & 60069.92 & 640 & 64 & T60 \\
    2023--05--07 & 60071.02 & 150 & 15 & T60 \\
    2023--05--08 & 60072.04 & 640 & 64 & T60 \\
    2023--05--08 & 60072.94 & 640 & 64 & T60 \\
    2023--05--09 & 60073.92 & 640 & 64 & T60 \\
    2023--05--10 & 60074.91 & 640 & 64 & T60 \\
    2023--05--12 & 60076.90 & 640 & 64 & T60 \\
    2023--05--13 & 60077.90 & 640 & 64 & T60 \\
    2023--05--15 & 60079.03 & 580 & 58 & T60 \\
    2023--05--15 & 60079.89 & 640 & 64 & T60 \\
    \hline
    \\
    \\
    \\
    \\
    \end{tabular}
    \end{table}

    \begin{table}
   	\caption{Average polarization degrees ($P$), and polarization angles ($\theta$) of highly polarized stars.} 
   	\label{table:hp}
   	\centering
   	\renewcommand{\arraystretch}{1.0} 
   	\scalebox{1.0}{
   	\begin{tabularx}{0.49\textwidth}[c]{l c c c c} 
   		\hline\hline 
   		Star & Filter & $P$~[\%] & $\theta$~[deg] & Ref. \\ \hline
  		HD~204827  & $B$ & $5.789 \pm 0.011$ & $57.79 \pm 0.02$ & [1] \\ 
   		& $V$ & $5.602 \pm 0.019$ & $58.33 \pm 0.02$ & [1] \\  
            & $R$ & $5.079 \pm 0.011$ & $59.21 \pm 0.02$ & [1] \\
   		HD 25443 & $B$ & $5.232 \pm 0.092$ & $134.28 \pm 0.51$ & [2] \\
   		& $V$ & $5.127 \pm 0.061$ & $134.2 \pm 0.34$ & [2] \\
   		& $R$ & $4.734 \pm 0.045$ & $133.65 \pm 0.28$ & [2] \\

   		\hline
   	\end{tabularx}}
   	\tablebib{(1) \citet{2021AJ....161...20P}; (2) \citet{1992AJ....104.1563S}.}
   \end{table}

    \onecolumn

	\begin{longtable}{cccccccc}
	\caption{Numbers, Identifiers, coordinates, parallaxes, distance, reddening magnitudes, polarization degrees, polarization angles, and the number of polarimetric observations of the field stars.}
	\label{table:is} \\
	\hline
	\hline
	Number & Identifier & Coordinates & Parallax & Filter & $P$ & $\theta$ & $N_{\rm obs}$ \\
	& [Gaia DR3] & [J2000d] & [mas] & & [\%] & [deg] & \\
	\hline
	\endfirsthead
	\hline
	\hline
	Number & Identifier & Coordinates & Parallax & Filter & $P$ & $\theta$ & $N_{\rm obs}$ \\
	& [Gaia DR3] & [J2000d] & [mas] & & [\%] & [deg] & \\
	\hline
	\endhead

    &&&& $B$ & 0.4510$\pm$0.0714 & 63.6$\pm$4.5 & 48 \\
    1 & 4066065089814336768 & 271.3059076143, & 0.8970$\pm$0.0163 & $V$ & 
    0.4670$\pm$0.0505 & 67.1$\pm$3.1 & 48 \\
    && --24.3803046097 && $R$ & 0.5632$\pm$0.0433 & 68.9$\pm$2.2 & 48 \\

    &&&& $B$ & 2.3768$\pm$0.2326 & 60.8$\pm$2.8 & 20 \\
    2 & 4066053132629912320 & 271.3179724764, & 0.6584$\pm$0.0428 & $V$ & 
    2.2920$\pm$0.1523 & 60.0$\pm$1.9 & 20 \\
    && --24.4198101995 && $R$ & 2.3349$\pm$0.0878 & 61.1$\pm$1.1 & 20 \\

    &&&& $B$ & 0.5098$\pm$0.0216 & 50.6$\pm$1.2 & 15 \\
    3 & 4066051522042471680 & 271.2812461348, & 0.8377$\pm$0.0202 & $V$ & 
    0.3841$\pm$0.0353 & 53.9$\pm$2.6 & 15 \\
    && --24.4997570914 && $R$ & 0.4584$\pm$0.0173 & 53.0$\pm$1.1 & 15 \\

    &&&& $B$ & 0.5116$\pm$0.0135 & 9.3$\pm$0.8 & 16 \\
    4 & 4066052617233631488 & 271.2459306023, & 0.8165$\pm$0.0209 & $V$ & 
    0.5845$\pm$0.0208  & 8.4$\pm$1.0 & 16 \\
    && --24.4564830252 && $R$ & 0.5502$\pm$0.0188 & 6.0$\pm$1.0 & 16 \\

    &&&& $B$ & 0.3849$\pm$0.0668 & 65.7$\pm$4.9 & 16 \\
    5 & 4066051281524301696 & 271.3092838328, & 0.8910$\pm$0.0145 & $V$ & 
    0.2134$\pm$0.0565 & 57.4$\pm$7.4 & 16 \\
    && --24.5108898514 && $R$ & 0.2802$\pm$0.0298 & 54.7$\pm$3.0 & 16 \\

    &&&& $B$ & 0.6153$\pm$0.0410 & 60.7$\pm$1.9 & 16 \\
    6 & 4066064540063670016 & 271.2674438044, & 0.8975$\pm$0.0155 & $V$ & 
    0.6890$\pm$0.0440 & 58.2$\pm$1.8 & 16 \\
    && --24.4193396203 && $R$ & 0.6409$\pm$0.0289 & 66.8$\pm$1.3 & 16 \\

    &&&& $B$ & 1.0775$\pm$0.0297 & 172.3$\pm$0.8 & 14 \\
    7 & 4066064819261901440 & 271.2344044218, & 0.7160$\pm$0.0171 & $V$ & 
    0.7497$\pm$0.0473 & 175.1$\pm$1.8 & 14 \\
    && --24.4010174976 && $R$ & 1.2141$\pm$0.0249 & 172.3$\pm$0.6 & 14 \\

    &&&& $B$ & 0.1635$\pm$0.1177 & 5.9$\pm$17.9 & 15 \\
    8 & 4066053201350041088 & 271.2972066384, & 0.8567$\pm$0.0348 & $V$ & 
    0.3246$\pm$0.0435 & 14.8$\pm$3.8 & 15 \\
    && --24.4103922466 && $R$ & 0.3290$\pm$0.0466 & 7.7$\pm$4.0 & 15 \\

    &&&& $B$ & 0.3899$\pm$0.1245 & 64.7$\pm$8.9 & 16 \\
    9 & 4066052995190959360 & 271.3533088425, & 0.8305$\pm$0.0245 & $V$ & 
    1.1575$\pm$0.2386 & 46.8$\pm$5.8 & 16 \\
    && --24.4168530036 && $R$ & 0.9917$\pm$0.1533 & 48.4$\pm$4.4 & 16 \\

    &&&& $B$ & 0.5685$\pm$0.0469 & 58.6$\pm$2.4 & 16 \\
    10 & 4066055095455592576 & 271.3894338404, & 0.8200$\pm$0.0221 & $V$ & 
    0.5259$\pm$0.0372 & 65.5$\pm$2.0 & 16 \\
    && --24.3548659296 && $R$ & 0.4354$\pm$0.0186 & 61.9$\pm$1.2 & 16 \\

    &&&& $B$ & 0.3561$\pm$0.0558 & 14.5$\pm$4.5 & 16 \\
    11 & 4066066979600006656 & 271.3191660725, & 0.8094$\pm$0.0163 & $V$ & 
    0.3576$\pm$0.0446 & 179.8$\pm$3.6 & 16 \\
    && --24.3315901481 && $R$ & 0.4983$\pm$0.0391 & 11.7$\pm$2.2 & 16 \\

    &&&& $B$ & 0.4807$\pm$0.0273 & 66.7$\pm$1.6 & 16 \\
    12 & 4066066498563609344 & 271.2211466119, & 0.6556$\pm$0.0366  & $V$ & 
    0.4835$\pm$0.0299 & 65.8$\pm$1.8 & 16 \\
    && --24.3244706110 && $R$ & 0.5850$\pm$0.0229 & 66.7$\pm$1.1 & 16 \\

    &&&& $B$ & 2.4666$\pm$0.0888 & 34.2$\pm$1.0 & 24 \\
    13 & 4066053755426343936 & 271.3864322875, & 0.6348$\pm$0.0190 & $V$ & 
    2.5907$\pm$0.0931 & 30.7$\pm$1.0 & 24 \\
    && --24.4088938655 && $R$ & 2.3911$\pm$0.0586 & 29.3$\pm$0.7 & 24 \\

    &&&& $B$ & 0.6839$\pm$0.1104 & 57.0$\pm$4.6 & 16 \\
    14 & 4066066429849761408 & 271.2238606428, & 0.8021$\pm$0.0155 & $V$ & 
    0.9438$\pm$0.0612 & 61.3$\pm$1.9 & 16 \\
    && --24.3248992994 && $R$ & 0.8156$\pm$0.0709 & 63.1$\pm$2.5 & 16 \\

    &&&& $B$ & 0.6114$\pm$0.1421 & 6.5$\pm$6.5 & 16 \\
    15 & 4066064814936438400 & 271.2417881244, & 0.6839$\pm$0.0178 & $V$ & 
    0.6839$\pm$0.0652 & 4.9$\pm$2.7 & 16 \\
    && --24.4057311137 && $R$ & 0.6443$\pm$0.0272 & 6.4$\pm$1.2 & 16 \\

    &&&& $B$ & 2.7302$\pm$0.1238 & 2.0$\pm$1.3 & 14 \\
    16 & 4066052209237880576 & 271.3439872723, & 0.9004$\pm$0.0164 & $V$ &   
    2.8448$\pm$0.0340 & 174.7$\pm$0.3 & 14 \\
    && --24.4397975463 && $R$ & 3.0119$\pm$0.0259 & 175.6$\pm$0.2 & 14 \\

    &&&& $B$ & 0.2941$\pm$0.0225 & 75.6$\pm$2.2 & 19 \\
    17 & 4065970398702755968 & 271.1991211204, & 0.8158$\pm$0.0181 & $V$ & 
    0.2673$\pm$0.0344 & 75.2$\pm$3.7 & 19 \\
    && --24.4733855530 && $R$ & 0.1920$\pm$0.0232 & 78.7$\pm$3.4 & 19 \\

    &&&& $B$ & 2.3064$\pm$0.0343 & 159.9$\pm$0.4 & 24 \\
    18 & 4066064643142852224 & 271.2108418248, & 0.7632$\pm$0.0153 & $V$ & 
    2.6995$\pm$0.0497 & 159.3$\pm$0.5 & 24 \\
    && --24.4283694408 && $R$ & 2.7587$\pm$0.0269 & 159.0$\pm$0.3 & 24 \\

    &&&& $B$ & 4.1504$\pm$0.1093 & 13.7$\pm$0.8 & 24 \\
    19 & 4066066120606487552 & 271.2389172227, & 0.5473$\pm$0.0307 & $V$ & 
    3.5982$\pm$0.0123  & 6.0$\pm$0.1 & 24 \\
    && --24.3468236972 && $R$ & 3.9686$\pm$0.0096 & 9.7$\pm$0.1 & 24 \\

    &&&& $B$ & 0.9895$\pm$0.2411 & 62.2$\pm$6.8 & 20 \\
    20 & 4066052793353449856 & 271.2741311882, & 0.8720$\pm$0.0195 & $V$ & 
    0.3237$\pm$0.1837 & 71.2$\pm$14.8 & 20 \\
    && --24.4239033454 && $R$ & 0.5309$\pm$0.0684 & 72.4$\pm$3.7 & 20 \\

    &&&& $B$ & 0.6156$\pm$0.1215 & 58.4$\pm$5.6 & 20 \\
    21 & 4066052758993708032 & 271.2639480402, & 0.8527$\pm$0.0207 & $V$ & 
    0.2688$\pm$0.1223 & 49.9$\pm$12.2 & 20 \\
    && --24.4454711067 && $R$ & 0.6073$\pm$0.1010 & 54.4$\pm$4.7 & 20 \\
    
    &&&& $B$ & 2.3927$\pm$0.4187 & 161.2$\pm$5.0 & 20 \\
    22 & 4066065712615114880 & 271.2143725965, & 0.6601$\pm$0.0266 & $V$ & 
    1.7056$\pm$0.1441 & 162.2$\pm$2.4 & 20 \\
    && --24.3675817201 && $R$ & 2.5474$\pm$0.0729 & 166.2$\pm$0.8 & 20 \\

    &&&& $B$ & 0.0698$\pm$0.0579 & 174.6$\pm$19.8 & 20 \\
    23 & 4066066743407427328 & 271.3469938928, & 0.8506$\pm$0.0175 & $V$ & 
    0.4647$\pm$0.0979 & 33.1$\pm$5.9 & 20 \\
    && --24.3345472232 && $R$ & 1.1782$\pm$0.1597 & 169.7$\pm$3.9 & 20 \\

    &&&& $B$ & 2.0197$\pm$0.3458 & 132.1$\pm$4.9 & 16 \\
    24 & 4066052862072911104 & 271.3368141244, & 0.8389$\pm$0.0244 & $V$ & 
    2.8932$\pm$0.2451 & 39.4$\pm$2.4 & 16 \\
    && --24.4427979636 && $R$ & 2.8648$\pm$0.1942 & 47.7$\pm$1.9 & 16 \\

    \hline	
    \end{longtable}

    \section{Formulae}\label{sec:drissen}

    The formulae given by \citet{1986ApJ...304..188D} are given in this section. For the orbital inclination ($i$):

    \begin{equation}
    \begin{split}
    \begin{aligned}
    \left(\frac{1 - \cos i}{1 + \cos i}\right)^4 
    & =  \frac{(u_1 + q_2)^2 + (u_2 - q_1)^2}{(u_2 + q_1)^2 + (u_1 - q_2)^2} \\
    & =  \frac{(u_3 + q_4)^2 + (u_4 - q_3)^2}{(u_4 + q_3)^2 + (u_3 - q_4)^2}.		
    \label{Inclination}
    \end{aligned}
    \end{split}
    \end{equation}
    
    Longitude of ascending node ($\Omega$) can be computed using the following:
    	
    \begin{equation}
    \tan \Omega = \frac{A+B}{C+D} = \frac{C-D}{A-B},
    \label{Orientation}
    \end{equation}

    where,
	
    \begin{equation}
    \begin{split}
    A = \frac{u_4 - q_3}{(1 - \cos i)^2}, \:\:\:\:\:\:\:\:\:\: B = \frac{u_4 + q_3}{(1 + \cos i)^2}, \\
    C = \frac{q_4 - u_3}{(1 + \cos i)^2}, \:\:\:\:\:\:\:\:\:\: D = \frac{u_3 + q_4}{(1 - \cos i)^2}.
    \end{split}
    \label{alphabets1}
    \end{equation}

    The following formulae can be used to derive $A_{\rm q}$ and $A_{\rm u}$:

    \begin{equation}
    \begin{split}
    A_{\rm q} = \left(\frac{q_3^2 + q_4^2}{q_1^2 + q_2^2}\right)^{\frac{1}{2}}\left(\frac{J^2 + K^2}{N^2+R^2}\right)^{\frac{1}{2}} , \:\:\:\:\: A_{\rm u} = \left(\frac{u_3^2 + u_4^2}{u_1^2 + u_2^2}\right)^{\frac{1}{2}} \left(\frac{E^2 + F^2}{L^2+M^2}\right)^{\frac{1}{2}},    
    \end{split}
    \label{Ratios}
    \end{equation}
    
    where,

    \begin{equation}
    \begin{split}
    E &= \sin 2i \cos \Omega, \:\:\:\:\:\:\:\:\:\:\:\: F = 2 \sin i \sin \Omega, \:\:\:\:\:\:\:\:\:\:\:\: J = \sin 2i \sin \Omega,  \:\:\:\:\:\:\:\:\:\:\:\: K = 2 \sin i \cos \Omega, \\
    L &= (1 + \cos^2 i) \cos \Omega, \:\:\:\:\:\:\:\:\:\:\:\: M = 2 \cos i \sin \Omega, \:\:\:\:\:\:\:\:\:\:\:\: N = (1 + \cos^2 i) \sin \Omega,  \:\:\:\:\:\:\:\:\:\:\:\: R = 2 \cos i \cos \Omega. 
    \end{split}
    \label{alphabets2}
    \end{equation}

    We can computer the moment $\tau_0$H as follows:

    \begin{equation}
    \begin{split}
    \tau_0 H = \left(\frac{q_3^2 + q_4^2}{L^2 + M^2}\right)^{\frac{1}{2}} = \left(\frac{u_3^2 + u_4^2}{N^2 + R^2}\right)^{\frac{1}{2}}.    
    \end{split}
    \label{Moment}
    \end{equation}

    \end{document}